\documentclass[aps,prd,reprint,superscriptaddress,nofootinbib,floatfix]{revtex4-2}

\usepackage{amsmath,amssymb,amsthm}
\usepackage{graphicx}
\usepackage[hidelinks]{hyperref}
\newtheorem{proposition}{Proposition}

\newcommand{\Om}{\Omega_{\rm m}}
\newcommand{\ODE}{\Omega_{\rm DE}}
\newcommand{\rDE}{\rho_{\rm DE}}
\newcommand{\rhom}{\rho_{\rm m}}
\newcommand{\mpl}{M_{\rm P}}

\begin{document}

\title{Stealth scalars on prescribed dark-energy backgrounds:\\ reconstruction, branch regularity, and perturbative bounds}

\author{Jos\'e Blanco}
\email{jose.blancova@postgrado.uv.cl}
\affiliation{Instituto de F\'isica y Astronom\'ia,
Universidad de Valpara\'iso,
Avenida Gran Breta\~na 1111, Valpara\'iso, Chile}
\affiliation{Facultad de Matem\'aticas, Universidad Veracruzana,
91097 Xalapa, Veracruz, Mexico}

\author{Cuauhtemoc Campuzano}
\email{ccampuzano@uv.mx}
\affiliation{Facultad de F\'isica, Universidad Veracruzana,
91097 Xalapa, Veracruz, Mexico}

\author{V\'ictor H. C\'ardenas}
\affiliation{Instituto de F\'isica y Astronom\'ia,
 Universidad de Valpara\'iso,
Avenida Gran Breta\~na 1111, Valpara\'iso, Chile}

\author{Giovany Cruz}
\affiliation{Facultad de F\'isica, Universidad Veracruzana,
91097 Xalapa, Veracruz, Mexico}

\begin{abstract}
We reconstruct homogeneous stealth scalar fields on prescribed Friedmann--Lema\^itre--Robertson--Walker backgrounds with constant or evolving dark-energy equations of state. The scalar stress tensor vanishes, so the expansion history is supported by an independent background sector. Eliminating the self-interaction potential yields a Riccati equation for the scalar profile. We use it to distinguish exact stealthness from global regularity and from the stronger identification of a term in the scalar stress tensor with the prescribed dark-energy density. We prove that this sector-preserving identification is not dynamically invariant in a mixed flat constant-$w$ cosmology. For radiation-free $w$CDM backgrounds, a hypergeometric representation allows the regular and singular branches to be separated by a background-dependent threshold. At conformal coupling the reconstruction instead reduces to an exact comoving-distance quadrature, valid for arbitrary regular flat histories, including radiation. A CPL phantom-divide crossing does not singularize the background stealth equation, although the perturbations of the prescribed dark-energy sector require an independent treatment. The scalar--tensor description gives luminal high-frequency propagation under a positive effective Planck mass. If the field tracks its cosmological value locally without screening, Cassini constraints strongly suppress the corresponding late-time gravitational responses.
\end{abstract}

\maketitle

% ============================================================================
\section{Introduction}\label{sec:intro}

The acceleration of the cosmic expansion was established through Type~Ia supernovae~\cite{Riess1998,Perlmutter1999}, and the $\Lambda$CDM description remains consistent with precision CMB observations~\cite{Planck2018}. The physical origin of the accelerated expansion, however, is not settled. Alternatives include dynamical dark-energy fields and effective fluids~\cite{Frieman2008,Copeland2006,Wetterich1988,Ratra1988,Caldwell1998}. Constant-$w$ and Chevallier--Polarski--Linder (CPL) parametrizations~\cite{Chevallier2001,Linder2003} supply useful expansion histories, particularly in view of recent DESI BAO constraints~\cite{DESI2025}. In this work these histories are prescribed, rather than generated by the additional scalar field.

A \emph{stealth scalar} is a nontrivial scalar configuration on a given geometry for which the full scalar contribution to the gravitational field equations vanishes, $T^{(S)}_{\mu\nu}=0$, even though the scalar is not constant. Its kinetic, potential, and nonminimal-curvature terms need not vanish separately; rather, they cancel in the total energy--momentum tensor. The cancellation is possible because varying the curvature coupling contributes terms of the type associated with the improved stress tensor~\cite{Callan1970}. Early explicit realizations included time-dependent configurations on a $(2+1)$-dimensional black-hole background~\cite{Ayon2006} and nonminimally coupled fields on flat spacetime~\cite{Ayon2005}. Further examples were found on rotating three-dimensional black holes~\cite{Hassaine2014}, while stealth fields were also considered in semiclassical cosmological creation~\cite{Maeda2012}. The defining property is therefore \emph{absence of background backreaction}, not absence of scalar dynamics.

The cosmological literature makes this distinction especially important. Ay\'on-Beato \emph{et al.} proved that standard FLRW cosmologies admit conformal stealth configurations, including inhomogeneous fields compatible with a homogeneous and isotropic geometry~\cite{Ayon2013}. Their later nonconformal construction examined radiation- and matter-dominated cosmological histories~\cite{Ayon2018}. Perfect-fluid and dust backgrounds were investigated by Alvarez \emph{et al.}~\cite{Alvarez2019}, whereas Campuzano \emph{et al.}~\cite{Campuzano2016} proposed a $\Lambda$CDM mimicry construction that attributes cosmological significance to scalar-sector contributions. These studies establish that both the Riccati-type reduction and the conformal reconstruction have precedents; the present question instead concerns the compatibility of a sectorwise identification with the \emph{full} stealth evolution on an independently prescribed background.

The way stealth conditions are imposed also matters. In a reverse-construction approach, Erices \emph{et al.} obtained the geometries compatible with a specified stealth ansatz and analyzed thermodynamic consequences~\cite{Erices2025}. The present calculation adopts the complementary, geometry-first problem: the FLRW expansion history $H(z)$ is fixed in advance, and we determine the homogeneous scalar field and its self-interaction. Extensions of stealth configurations within more general scalar--tensor transformations have also been discussed~\cite{FaraoniZeyn2024}. These frameworks are not assumed here: our action contains a canonical kinetic term and a nonminimal $\zeta R\phi^2$ coupling.

Vanishing \emph{background} stress energy does not imply that a stealth field is physically featureless. Thermodynamic imprints have been found in black-hole settings~\cite{Bakopoulos2025,Erices2025}; stealth configurations have also been discussed as metastable or thermally non-equilibrated states in scalar--tensor thermodynamics~\cite{FaraoniFranconnet2022,Giardino2023}. Their perturbative behavior is a distinct question: gauge-invariant stability tests on flat scalar--tensor backgrounds reveal both stable and unstable cases~\cite{Faraoni2010}, while a restricted non-geometrical perturbation analysis has been carried out for homogeneous stealth dust~\cite{Blanco2026}. None of these calculations establishes the coupled, gauge-invariant stability of the prescribed $w$CDM/CPL solutions considered here. This motivates treating background existence, branch regularity, and perturbative viability as logically separate requirements. Our spectator scalar should also be distinguished from nonminimally coupled quintessence models in which the scalar contributes to cosmic acceleration~\cite{Wolf2025,PanYe2026}.

For definiteness, the prescribed CPL history is written as
\begin{equation}
 w(a)=w_0+w_a(1-a).
 \label{eq:CPL}
\end{equation}
The associated dark-energy density belongs to the external background sector, not to the stealth scalar. Interpreting the reconstructed scalar-potential term as the independently prescribed dark-energy density is therefore an additional condition, stronger than the exact stealth requirement; its precise algebraic form is introduced in Sec.~\ref{sec:mimicry}. As anticipated by the sector-degeneracy issue emphasized in Ref.~\cite{Kunz2009}, agreement at the level of total gravitational sourcing does not justify attributing the same dynamics to individual contributions. We test this distinction directly: in mixed flat constant-$w$ cosmologies the sector-preserving surface is not invariant under the reconstructed stealth evolution.

The remaining analysis treats three logically separate questions. First, the full stealth equations reconstruct $\phi(z)$ and $V(\phi)$ on the imposed geometry. Second, the Riccati flow identifies regular branches and their background-dependent separatrix for radiation-free $w$CDM; at conformal coupling, an equivalent comoving-distance quadrature gives a pole criterion that also accommodates radiation, consistently with the earlier conformal construction~\cite{Ayon2013}. Third, we examine the CPL background and the restricted high-frequency and quasistatic predictions of the corresponding scalar--tensor theory. In this last step the effective Planck mass and local bounds must be considered even when $T^{(S)}_{\mu\nu}=0$ on the background~\cite{Clifton2012,Joyce2015}. We do not claim that the stealth field explains the prescribed dark energy, that its potential is natural, or that background regularity establishes full perturbative stability.

\section{Stealth reconstruction and regularity}\label{sec:recon}
% ============================================================================

We take the background sources to be independent of a canonical scalar field with nonminimal curvature coupling~\cite{Campuzano2016}:
\begin{align}
S=\int d^4x\sqrt{-g}\Bigl[&\frac{R}{2\kappa}+\mathcal L_{\rm bg}
-\frac12(\nabla\phi)^2\nonumber\\
&-\frac12\zeta R\phi^2-V(\phi)\Bigr].
\label{eq:action}
\end{align}
Here $\kappa=8\pi G$ and $\zeta\neq0$, while $\mathcal L_{\rm bg}$ contains the sources of the prescribed geometry: pressureless matter, radiation where included, and an effective dark-energy component. The additional scalar contribution to the Einstein equations is
\begin{align}
T^{(S)}_{\mu\nu}={}&
\nabla_\mu\phi\nabla_\nu\phi
-g_{\mu\nu}\left[V+\frac12(\nabla\phi)^2\right]
\nonumber\\
&+\zeta\left(G_{\mu\nu}\phi^2
-\nabla_\mu\nabla_\nu\phi^2
+g_{\mu\nu}\Box\phi^2\right).
\label{eq:Tmn}
\end{align}
For a nonconstant field with $T^{(S)}_{\mu\nu}=0$, the Einstein equations reduce to those of the prescribed background sector. The contracted Bianchi identity gives
\[
\nabla^\mu T^{(S)}_{\mu\nu}
=(\Box\phi-\zeta R\phi-V_{,\phi})\nabla_\nu\phi
\]
wherever $\nabla_\mu\phi\neq0$. The minimally coupled canonical field has no nonconstant homogeneous stealth solution of this type on a regular FLRW background; hence the nonminimal term is essential.

\subsection{Geometry-first reconstruction}

For a homogeneous scalar $\phi(t)$ on an FLRW background, define
\[
 u\equiv\frac{\dot\phi}{\phi},
 \qquad
 Q\equiv-\frac{V}{\zeta\phi^2}.
\]
The temporal stealth equation fixes
\begin{equation}
Q=3\left(H^2+\frac{k}{a^2}\right)+6Hu+\frac{u^2}{2\zeta},
\label{eq:Q}
\end{equation}
while eliminating $V$ between the temporal and spatial components gives
\begin{equation}
\dot u-Hu+\gamma u^2+\dot H-\frac{k}{a^2}=0,
\qquad
\gamma\equiv2-\frac{1}{2\zeta}.
\label{eq:master}
\end{equation}
Appendix~\ref{app:flrw} supplies the reduction. Closely related reconstruction equations were derived for nonconformal cosmological stealths~\cite{Ayon2018}. Here Eq.~\eqref{eq:master} is used to reconstruct the field on an independently specified background and to analyze the global flow without imposing a sectorwise identification of the potential.

Using redshift, we introduce
\begin{align*}
 X&\equiv(1+z)\frac{d\ln\phi}{dz}=-\frac{u}{H},\\
 A&\equiv(1+z)\frac{H'}{H},\qquad
 \Omega_k\equiv-\frac{k}{a^2H^2}.
\end{align*}
Then
\begin{equation}
(1+z)X'=A-\Omega_k-(1+A)X-\gamma X^2.
\label{eq:masterz}
\end{equation}
All dependence on the target cosmology is carried by $A(z)$ and, if present, $\Omega_k(z)$. Once $X$ is known,
\begin{align}
\phi(z)&=\phi_0\exp\!\left[\int_0^z\frac{X(\tilde z)}{1+\tilde z}\,d\tilde z\right],
\nonumber\\
V(z)&=-3\zeta H^2\phi^2\left(1-\Omega_k-2X+\frac{X^2}{6\zeta}\right).
\label{eq:reconstruction}
\end{align}
The initial datum for $X$ selects a branch; integrating it determines $\phi(z)$ and $V(z)$. A monotonic field profile ensures that this parametric curve assigns a unique $V$ to each field value reached by the solution.

For $\gamma\neq0$, the Riccati equation can be linearized with
\begin{align}
 \Psi&\equiv\left(\frac{\phi}{\phi_0}\right)^\gamma,
 \qquad X=\frac{1+z}{\gamma}\frac{\Psi'}{\Psi},\nonumber\\
 \ddot\Psi&-H\dot\Psi
 +\gamma\left(\dot H-\frac{k}{a^2}\right)\Psi=0.
 \label{eq:linear}
\end{align}
A zero of $\Psi$ therefore signals a pole of $X$. The special case $\zeta=1/4$ ($\gamma=0$) is regular: Eq.~\eqref{eq:master} is then already linear in $u$.

\subsection{Monotonicity and regularity}\label{sec:theorems}

At $X=0$, Eq.~\eqref{eq:masterz} gives
\begin{equation}
(1+z)X'\big|_{X=0}
=-\frac{\dot H-k/a^2}{H^2}
=\frac{\kappa(\rho_{\rm tot}+p_{\rm tot})}{2H^2}.
\label{eq:barrier}
\end{equation}
With the strict null energy condition (NEC), the vector field points toward positive $X$ at the axis. More precisely, let $X(z)$ be regular on $[z_0,z_1]$, with $X(z_0)>0$, and suppose $\rho_{\rm tot}+p_{\rm tot}>0$ throughout this interval. Then $X(z)>0$ for every $z\in[z_0,z_1]$. To see this, a first zero $z_*$ approached from positive values would require $X'(z_*)\le0$, while Eq.~\eqref{eq:barrier} instead implies $X'(z_*)>0$. Thus $X$ cannot vanish. Since $d\phi/dz=X\phi/(1+z)$ and Eq.~\eqref{eq:reconstruction} preserves the sign of $\phi_0$, the field is strictly monotonic, so the parametric reconstruction determines a single-valued $V(\phi)$ over the corresponding field interval.

The positive branch also admits a comparison bound when $\gamma\ge0$, equivalently $\zeta<0$ or $\zeta\ge1/4$. Assume the NEC on $[0,z_1]$ and $X_0\equiv X(0)>0$. Then
\begin{equation}
0<X(z)\le X_{\rm lin}(z),\quad
X_{\rm lin}(z)=\frac{X_0+I(z)}{(1+z)E(z)},
\label{eq:Xlin}
\end{equation}
where $E=H/H_0$ and
\begin{equation}
 I(z)=\int_0^z E(\tilde z)
 [A(\tilde z)-\Omega_k(\tilde z)]\,d\tilde z.
\end{equation}
Indeed, for $\gamma\ge0$ and $X>0$, the term $-\gamma X^2$ in Eq.~\eqref{eq:masterz} is nonpositive, so the Riccati flow is bounded above by the corresponding linear equation. Equation~\eqref{eq:Xlin} solves that equation with $X_{\rm lin}(0)=X_0$. The comparison theorem yields $X\le X_{\rm lin}$, and the positivity argument above supplies the strict lower bound. In particular, no positive pole can occur on the interval while the prescribed background remains regular.

For a flat matter--dark-energy background the NEC becomes $1+w(z)\ODE(z)>0$, which allows part of the phantom regime and smooth crossings of $w=-1$. The complementary range $0<\zeta<1/4$ has $\gamma<0$ and can develop positive poles. Near a pole, with $q=\ln(1+z)$,
\[
X\simeq\frac{1}{|\gamma|(q_s-q)},
\qquad
\phi\propto(q_s-q)^{-1/|\gamma|},
\]
so the Riccati divergence corresponds to a genuine scalar-field singularity. The same conclusion does not hold with the same field asymptotics for $\gamma>0$: a negative-$X$ pole corresponds to a zero of $\Psi=(\phi/\phi_0)^\gamma$, hence $\phi\to0$, rather than $|\phi|\to\infty$. For $\gamma>1$, $|\dot\phi|$ still diverges as the zero is approached, since $\phi\propto(q_s-q)^{1/\gamma}$. Thus a pole of $X$ need not have the same manifestation in $\phi$ on different branches.

% ============================================================================
\section{Geometrical stealth and sector-preserving mimicry}\label{sec:mimicry}
% ============================================================================

The reconstruction above uses only the target geometry. To compare it with constructions in which the potential is identified directly with the dark-energy sector, write
\[
Q=\kappa\rDE+\Delta,
\qquad
\delta\equiv\frac{\Delta}{3H^2}.
\]
For a flat matter--dark-energy background, Eq.~\eqref{eq:Q} gives
\begin{equation}
\delta=\Om-2X+\frac{X^2}{6\zeta},
\qquad
V=-3\zeta H^2\phi^2(\ODE+\delta).
\label{eq:delta}
\end{equation}
The dimensionless mismatch $\delta$ compares the potential contribution with the chosen dark-energy density; it has no separate stress tensor.

We call $\delta=0$, or equivalently $-V/(\zeta\phi^2)=\kappa\rDE$, \emph{sector-preserving mimicry}. Under this additional condition,
\begin{align*}
\kappa\rhom&=-6Hu-\frac{u^2}{2\zeta},\\
\dot u+2Hu+\beta u^2&=\frac{\kappa}{2}(1+w)\rDE,
\qquad
\beta\equiv\frac{8\zeta-1}{4\zeta}.
\end{align*}
For $w=-1$ the second equation integrates to $a^2\phi^\beta u=\mathrm{const.}$, an intermediate relation also used in the $\Lambda$CDM mimicry construction of Ref.~\cite{Campuzano2016}. There the additional identification $-V/(\zeta\phi^2)=\Lambda$ is imposed and the expansion rate is reconstructed from the resulting scalar--matter relations. Our question is different: we \emph{independently prescribe} the standard $\Lambda$CDM history, require the full stealth equations, and then test whether the identification remains true throughout that history. The shared intermediate equation does not establish the dynamical consistency of the extra identification. In particular, Proposition~\ref{prop:noinv} also covers $w=-1$ during a mixed dust--$\Lambda$ epoch; it should not be read as a claim that Ref.~\cite{Campuzano2016} performed this invariance test.

Instantaneously, $\delta=0$ defines the algebraic branches
\begin{equation}
X_\pm=6\zeta\pm\sqrt{36\zeta^2-6\zeta\Om}.
\label{eq:Xpm}
\end{equation}
Equation~\eqref{eq:Xpm} specifies instantaneous compatibility, not a solution of the differential equation~\eqref{eq:masterz}.

For a flat constant-$w$ background, the flow normal to the surface $\delta=0$ is
\begin{equation}
\left.\frac{d\delta}{d\ln a}\right|_{\delta=0}
=\frac{\mathcal N_{w,\zeta}(X)}{12\zeta^2},
\label{eq:flow}
\end{equation}
with
\begin{align}
\mathcal N_{w,\zeta}(X)={}&
2(X-6\zeta)\left[(4\zeta-1)X^2+5\zeta X-3\zeta\right]
\nonumber\\
&+w\left[X^2-12\zeta X+6\zeta\right]
\left[(6\zeta-1)X+6\zeta\right].
\label{eq:Npoly}
\end{align}
Appendix~\ref{app:compat} gives the algebra leading to this expression.

\begin{proposition}[Non-invariance of sector-preserving mimicry]\label{prop:noinv}
For a mixed flat constant-$w$ cosmology with $\rhom\neq0$, $\rDE\neq0$, $w\neq0$, and $\zeta\neq0$, the surface $\delta=0$ is not an invariant submanifold of the exact stealth dynamics.
\end{proposition}

\begin{proof}
For constant $w$, $\rDE/\rhom\propto a^{-3w}$ varies continuously through a mixed epoch. If $\delta=0$ were preserved, Eq.~\eqref{eq:Xpm} would force $X$ to vary on a continuous interval while satisfying $\mathcal N_{w,\zeta}(X)=0$. Hence the cubic would have to vanish identically. Its constant term, $36\zeta^2(1+w)$, requires $w=-1$, but then the coefficient linear in $X$ is $-24\zeta^2\neq0$. Therefore no such polynomial identity exists.
\end{proof}

The failure of invariance applies to the extra sectorwise identification, not to the stealth condition $T^{(S)}_{\mu\nu}=0$ itself. Imposing $\delta=0$ at one epoch selects an initial condition without guaranteeing that the trajectory remains on that surface. We therefore retain $\delta(z)$ as a diagnostic. Solutions with $\delta(0)=0$ will be called \emph{anchored}; in the radiation-free flat case their initial values are $X_0=X_{0,\pm}=X_\pm(\Omega_{\rm m0})$.

% ============================================================================
\section{$w$CDM backgrounds}\label{sec:wcdm}
% ============================================================================

For a flat background with pressureless matter and constant equation-of-state parameter $w$,
\begin{equation}
E^2(z)=\Omega_{\rm m0}(1+z)^3+(1-\Omega_{\rm m0})(1+z)^{3(1+w)},
\label{eq:wcdmE}
\end{equation}
and $A=\tfrac32(1+w\ODE)$. Equation~\eqref{eq:masterz} becomes
\begin{equation}
(1+z)X'=\frac32(1+w\ODE)
-\left(\frac52+\frac32w\ODE\right)X-\gamma X^2.
\label{eq:wcdmRiccati}
\end{equation}
The $\Lambda$CDM limit is recovered at $w=-1$.

\subsection{Domain of the pressureless-matter approximation}
\label{sec:radiation}

The constant-$w$ formulas above omit radiation. They describe an idealized matter--dark-energy system and should not be used unchanged at recombination or during radiation domination. The Riccati equation remains valid when radiation is included; the flat-background expansion law becomes
\begin{align}
 E^2(z)={}&\Omega_{\rm r0}(1+z)^4
 +\Omega_{\rm m0}(1+z)^3\nonumber\\
 &+\Omega_{\rm DE0}f_{\rm DE}(z),\nonumber\\
 \Omega_{\rm DE0}={}&1-\Omega_{\rm m0}-\Omega_{\rm r0}.
 \label{eq:Erad}
\end{align}
The factor $f_{\rm DE}$ is $(1+z)^{3(1+w)}$ at constant $w$, or the CPL density factor in Eq.~\eqref{eq:ECPL}. It follows that
\begin{align}
 2E^2 A={}&4\Omega_{\rm r0}(1+z)^4
 +3\Omega_{\rm m0}(1+z)^3\nonumber\\
 &+3[1+w(z)]\Omega_{\rm DE0}f_{\rm DE}(z).
 \label{eq:Arad}
\end{align}
In radiation domination $A\to2$, and the limiting Riccati flow becomes
\begin{equation}
 \frac{dX}{d\ln(1+z)}=2-3X-\gamma X^2.
 \label{eq:radflow}
\end{equation}
Its fixed points are not those of matter domination. In particular, when $\gamma\ne0$ and the discriminant is nonnegative, they satisfy
\begin{equation}
 X_{{\rm r},\pm}=\frac{-3\pm\sqrt{9+8\gamma}}{2\gamma}.
 \label{eq:radfixed}
\end{equation}
For $\gamma=0$ the finite fixed point is $2/3$. The matter-era fixed points and separatrix results below belong to the radiation-free model. With $\Omega_{\rm r0}>0$ they describe, at most, an intermediate matter-dominated regime. The early-time behavior of a generic nonconformal branch must instead be obtained from Eq.~\eqref{eq:masterz} using Eq.~\eqref{eq:Arad}.

\subsection{Analytical solution}

For $w\neq0$ and $\gamma\neq0$, introduce
\[
r\equiv\frac{\rDE}{\rhom}=r_0a^{-3w},
\qquad
r_0=\frac{1-\Omega_{\rm m0}}{\Omega_{\rm m0}}.
\]
The linear equation for $\Psi$ reduces to the Gauss hypergeometric equation (Appendix~\ref{app:hyper}), with generic solution
\begin{equation}
\Psi(r)=\sum_{\sigma=\pm}C_\sigma r^{s_\sigma}
{}_2F_1(a_\sigma,b_\sigma;c_\sigma;-r),
\label{eq:hyper}
\end{equation}
where
\begin{align*}
s_\pm&=\frac{-5\pm\sqrt{73-12/\zeta}}{12w},\\
c_\sigma&=1+\frac{5}{6w}+2s_\sigma,\\
a_\sigma+b_\sigma&=c_\sigma-\frac12,
\qquad
a_\sigma b_\sigma=-s_\sigma\left(ws_\sigma+\frac13\right).
\end{align*}
For $0<\zeta<12/73$, the discriminant is negative and $s_\pm$ form a complex-conjugate pair; their linear combination can still be chosen real. This is the Frobenius representation of the same real Riccati flow, not a complex physical scalar field. The constants follow from $\Psi(r_0)=1$ and $\Psi_{,r}(r_0)=\gamma X_0/(3wr_0)$. A zero of $\Psi$ gives a pole of $X$. For the numerical benchmark below, we evaluated the hypergeometric functions and their initial-condition derivatives with arbitrary-precision arithmetic, solved the resulting $2\times2$ linear system for $C_\pm$, and located the first positive-redshift zero by bracketing/root finding. The accompanying source archive includes the numerical cross-checks, implemented with SciPy and mpmath. For example, with $(w,\Omega_{\rm m0},\zeta)=(-1,0.3,0.175)$ and the anchored upper-branch datum $X_0=X_{0,+}$, the first positive-redshift zero occurs at $z_s=3.8168$; this agrees with the Riccati integration summarized in Sec.~\ref{sec:global}. Since the benchmark has $r_0>1$, the defining Gauss series does not converge at $-r_0$. The functions can be evaluated by analytic continuation, for example using the Pfaff identity ${}_2F_1(a,b;c;-r)=(1+r)^{-a}{}_2F_1(a,c-b;c;r/(1+r))$.

\subsection{Radiation-free matter regime and separatrix}

When $w<0$, the radiation-free model approaches matter domination as $z\to\infty$. In a cosmology with radiation, the same limit applies only within the matter-dominated epoch. With $q=\ln(1+z)$, the autonomous limiting equation is
\begin{equation}
\frac{dX}{dq}=\frac32-\frac52X-\gamma X^2.
\label{eq:matterflow}
\end{equation}
Its fixed points are
\begin{equation}
X_{{\rm m},\pm}=\frac{-5\pm\sqrt{73-12/\zeta}}{4\gamma}.
\label{eq:fixed}
\end{equation}
For positive coupling they are real only for
\begin{equation}
\zeta\ge\zeta_{\rm m}\equiv\frac{12}{73}\simeq0.16438,
\label{eq:critfixed}
\end{equation}
as already implied by the discriminant found in Ref.~\cite{Ayon2018}. We denote the attractive root by $X_{\rm m,s}=X_{{\rm m},+}$ and the repulsive root by $X_{\rm m,u}=X_{{\rm m},-}$.

For $\zeta<0$ or $\zeta\ge1/4$, the repulsive fixed point is nonpositive and positive trajectories belong to the basin of $X_{\rm m,s}$, consistently with the comparison bound in Sec.~\ref{sec:theorems}. For $\zeta_{\rm m}<\zeta<1/4$, both fixed points are positive and $X_{\rm m,u}$ separates regular trajectories from those running to a pole. When $0<\zeta<\zeta_{\rm m}$, the radiation-free matter flow has no real fixed points and positive branches can reach a finite-$q$ pole. For the anchored lower branch of the $\Lambda$CDM benchmark we find $z_s\simeq11$ and $22$ for $\zeta=0.13$ and $0.14$. A further pole at $z_s\simeq4.8\times10^3$ for $\zeta=0.16$ is a property only of the deliberately radiation-free model; at that redshift a physical background must include radiation and the pole position must be recomputed.

Within the radiation-free matter--dark-energy system, the repulsive matter fixed point continues into a separatrix $X_{\rm sep}(z)$, defined by $X_{\rm sep}\to X_{\rm m,u}$ as $z\to\infty$. Because regular solutions of the first-order equation cannot intersect, initial data below $X_{\rm sep}(0)$ approach the stable matter solution, whereas data above it develop a pole. For the anchored upper branch the transition occurs at
\begin{equation}
X_{0,+}(\zeta_{\rm sep})=X_{\rm sep}(0;\zeta_{\rm sep}).
\label{eq:zetacdef}
\end{equation}
For $\Omega_{\rm m0}=0.3$ we find $\zeta_{\rm sep}\simeq0.1810$ for $\Lambda$CDM, $0.1806$ for $w=-0.9$, and $0.1813$ for $w=-1.1$; the pure matter approximation gives $0.1725$. This transition is only weakly sensitive to the three late-time histories tested here and differs from the analytical matter-era threshold $\zeta_{\rm m}$. The threshold $\zeta_{\rm sep}$ depends on the full expansion history used to define the separatrix; its quoted value should not be transferred to a radiation-inclusive model.

The resulting past-directed regularity of anchored branches is summarized in Table~\ref{tab:regularity}.

\begin{table}[t]
\caption{Past-directed regularity in the radiation-free matter--dark-energy benchmark ($\Omega_{\rm m0}=0.3$), assuming the NEC. For $\zeta>0$, ``regular'' means no Riccati pole; $M_*^2>0$ must be checked separately. The table is not a classification of radiation-dominated evolution.}
\label{tab:regularity}
\begin{ruledtabular}
\begin{tabular}{lll}
Coupling & $\mathcal B_-$ & $\mathcal B_+$ \\
\hline
$\zeta<0$ & negative-$X$ pole & regular \\
$0<\zeta<\zeta_{\rm m}$ & pole & pole \\
$\zeta_{\rm m}\le\zeta<\zeta_{\rm sep}$ & regular & pole \\
$\zeta_{\rm sep}<\zeta<1/4$ & regular & regular \\
$\zeta\ge1/4$ & regular & regular
\end{tabular}
\end{ruledtabular}
\end{table}

% ============================================================================
\section{CPL backgrounds}\label{sec:cpl}
% ============================================================================

For the CPL equation of state
\[
w(z)=w_0+w_a\frac{z}{1+z},
\]
the flat expansion history is
\begin{align}
 E^2(z)={}&\Omega_{\rm m0}(1+z)^3\nonumber\\
 &+(1-\Omega_{\rm m0})(1+z)^{3(1+w_0+w_a)}
 \exp\!\left[-\frac{3w_a z}{1+z}\right].
 \label{eq:ECPL}
\end{align}
The reconstruction retains the same structure as Eq.~\eqref{eq:wcdmRiccati}, now with $w\to w(z)$ and the corresponding $\ODE(z)$. Although a generic hypergeometric reduction is lost, the local behavior is analytic. Writing
\[
X(z)=X_0+X_1z+\frac12X_2z^2+\mathcal O(z^3),
\]
one obtains
\begin{align}
X_1&=A_0-(1+A_0)X_0-\gamma X_0^2,
\nonumber\\
X_2&=A_0'(1-X_0)-(2+A_0+2\gamma X_0)X_1,
\label{eq:CPLseries}
\end{align}
with
\begin{align*}
A_0&=\frac32\left[1+w_0(1-\Omega_{\rm m0})\right],\\
A_0'&=\frac32(1-\Omega_{\rm m0})
\left(w_a+3w_0^2\Omega_{\rm m0}\right).
\end{align*}
These coefficients provide a local consistency check. The appearance of $w_a$ in $X_2$, rather than $X_1$, follows directly from the Taylor series of $w(z)$ near $z=0$.

Anchoring still fixes $X_{0,\pm}$ through Eq.~\eqref{eq:Xpm} with $\Om=\Omega_{\rm m0}$, independently of $w_0$ and $w_a$, and gives $V_0=-3\zeta H_0^2\phi_0^2(1-\Omega_{\rm m0})$. A phantom-divide crossing occurs at
\begin{equation}
z_\times=-\frac{1+w_0}{1+w_0+w_a},
\label{eq:zcross}
\end{equation}
when this value lies in the relevant redshift interval. No coefficient of the \emph{background stealth equation} is singular at $w=-1$, so the crossing does not, by itself, obstruct the scalar reconstruction. This is not a statement about the perturbative stability of a single dark-energy fluid crossing the phantom divide. A full Einstein--Boltzmann treatment of the prescribed CPL sector requires a consistent crossing prescription, such as PPF~\cite{FangHuLewis2008}, or a specified multicomponent completion. The positivity and comparison arguments of Sec.~\ref{sec:theorems} apply only where $1+w(z)\ODE(z)>0$ in the radiation-free model; with radiation, the corresponding total NEC includes $4\Omega_r/3$.

Strict sector preservation remains an additional constraint. The polynomial $\mathcal N_{w,\zeta}$ was obtained using the instantaneous flat-background identities $d\Omega_m/dN=3w(N)\Omega_m\Omega_{\rm DE}$ and $A=3[1+w(N)\Omega_{\rm DE}]/2$, which remain valid for time-dependent $w(N)$; no $w_{,N}$ term enters the first derivative of $\delta$. The algebraic flow therefore extends pointwise to CPL, although its derivative along the flow includes $w_{,N}$. At $z=0$ it requires $\mathcal N_{w_0,\zeta}(X_0)=0$, and preservation to the next order fixes
\begin{equation}
w_a=\left(\frac{\mathcal N_{,X}X_{,N}}{\mathcal N_{,w}}\right)_0,
\label{eq:CPLwaCompat}
\end{equation}
whenever the denominator is nonzero. Here $X_{,N}=dX/d\ln a=-X_1$ at $z=0$, so the displayed sign is equivalent to $w_a=-\mathcal N_{,X}X_1/\mathcal N_{,w}$. This condition applies only at points satisfying $\mathcal N=0$ with $\mathcal N_{,w}\neq0$. Under that additional requirement, $w_a$ is constrained rather than freely specified; no analogous restriction is imposed in the geometry-first reconstruction. If $w_0+w_a<0$ in the radiation-free model, the dark-energy term becomes subdominant and the matter-flow structure controls its formal high-redshift limit. When radiation is included, this statement applies only to the intermediate matter-dominated regime.

% ============================================================================
\section{Branches and reconstructed potentials}\label{sec:global}
% ============================================================================

\noindent\textit{Numerical implementation.} The numerical benchmarks and vector figures were obtained with Python. Riccati trajectories and quadratures were evaluated using SciPy, and mpmath was used for the hypergeometric functions requiring analytic continuation. The accompanying source archive contains the numerical implementations and checks.\par\smallskip

For the numerical illustrations we fix $\Omega_{\rm m0}=0.3$ and consider $\Lambda$CDM, $w=-0.9$, $w=-1.1$, and CPL with $(w_0,w_a)=(-0.9,-0.4)$. The CPL history crosses $w=-1$ at $z_\times=1/3$. These parameter values define benchmarks, not best fits to cosmological data.

\subsection{Anchored branches}

\begin{figure*}[t]
\includegraphics[width=\textwidth]{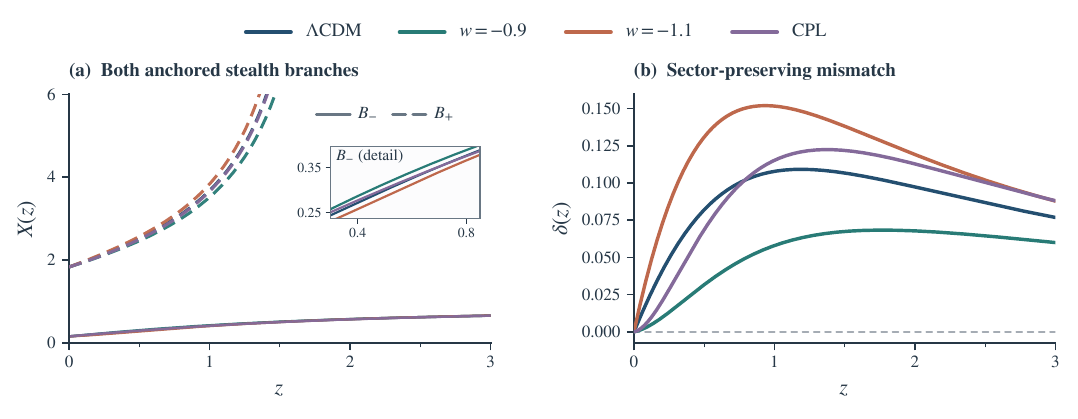}
\caption{Anchored conformal-stealth solutions for $\Omega_{\rm m0}=0.3$ and four background histories (colors). (a) Both trajectories: $B_-$ (solid) and $B_+$ (dashed). The inset magnifies $B_-$ at $0.30\leq z\leq0.85$ on the original $X(z)$ scale, without offsets. (b) Sector-preserving mismatch $\delta(z)$ along $B_-$. The curves are generated by numerical integration of the stated equations. The solutions satisfy $\delta(0)=0$ but generically depart from it at $z>0$.}
\label{fig:conformal}
\end{figure*}

\begin{figure*}[t]
\includegraphics[width=\textwidth]{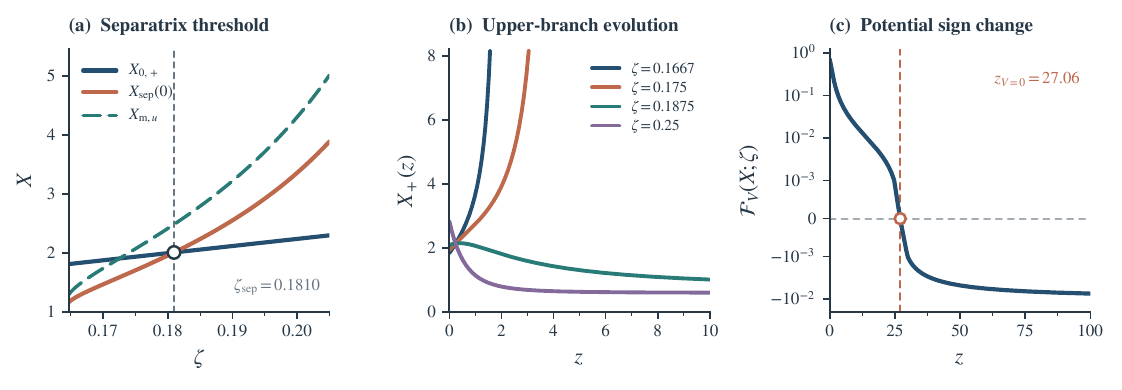}
\caption{Global structure of the idealized radiation-free $\Omega_{\rm m0}=0.3$ benchmark at positive coupling. (a) Anchored upper-branch value $X_{0,+}(\zeta)$, separatrix $X_{\rm sep}(0)$, and matter-era unstable fixed point $X_{\rm m,u}$ for the $\Lambda$CDM benchmark. The crossing gives $\zeta_{\rm sep}\simeq0.181$ for $\Omega_{\rm m0}=0.3$. (b) Upper-branch evolution on both sides of the transition. (c) Potential factor $\mathcal F_V(X;\zeta)$ along the lower branch for $\zeta=0.175$; its zero corresponds to a sign change of $V$ at $z\simeq27$. The curves follow from the specified radiation-free benchmark equations.}
\label{fig:basins}
\end{figure*}

At conformal coupling, $\zeta=1/6$, the anchored initial values are
\[
X_{0,\pm}=1\pm\sqrt{1-\Omega_{\rm m0}}.
\]
The two branches have sharply different global behavior. For all four backgrounds the lower branch $\mathcal B_-$ remains regular over the range displayed in Fig.~\ref{fig:conformal}, while $\delta(z)$ departs from zero immediately away from the anchoring epoch. The upper branch $\mathcal B_+$ develops a pole near $z\simeq2$, consistent with $1/6<\zeta_{\rm sep}$. Figure~\ref{fig:basins} shows the same transition away from conformal coupling: at $\zeta=0.175$ the upper branch diverges at $z_s\simeq3.82$, whereas the $\zeta=3/16$ and $1/4$ trajectories remain regular toward high redshift.

Representative values at $z=3$ are listed in Table~\ref{tab:benchmarks}. For positive coupling we show the lower branch, while for negative coupling we show the positive anchored branch.

\begin{table*}[t]
\caption{Representative anchored reconstructions at $z=3$ in the radiation-free benchmark with $\Omega_{\rm m0}=0.3$, with $\mathcal V\equiv V/(3H_0^2\phi_0^2)$. For $\zeta>0$ the last column gives the first pole of the upper branch when present; for $\zeta<0$ it gives the stable matter fixed point.}
\label{tab:benchmarks}
\begin{ruledtabular}
\begin{tabular}{llccccc}
$\zeta$ & Model & $X$ & $\phi/\phi_0$ & $\delta$ & $\mathcal V$ & $z_s(\mathcal B_+)$ or $X_{\rm m,s}$ \\
\hline
$1/6$ & $\Lambda$CDM & 0.6653 & 1.7844 & 0.0768 & $-1.1829$ & 1.958 \\
 & $w=-0.9$ & 0.6648 & 1.8032 & 0.0600 & $-1.2335$ & 2.032 \\
 & $w=-1.1$ & 0.6664 & 1.7683 & 0.0878 & $-1.1407$ & 1.896 \\
 & CPL & 0.6636 & 1.7845 & 0.0881 & $-1.1831$ & 1.964 \\
$3/16$ & $\Lambda$CDM & 0.6138 & 1.7383 & 0.0721 & $-1.2092$ & regular \\
$1/4$ & $\Lambda$CDM & 0.5403 & 1.6674 & 0.0788 & $-1.5766$ & regular \\
\hline
$-0.05$ & $\Lambda$CDM & 0.2577 & 1.3332 & 0.2280 & $+0.4655$ & 0.2644 \\
 & CPL & 0.2582 & 1.3330 & 0.2362 & $+0.4573$ & 0.2644 \\
$-1$ & $\Lambda$CDM & 0.4041 & 1.5182 & 0.1294 & $+7.5478$ & 0.4220 \\
 & $w=-0.9$ & 0.4013 & 1.5256 & 0.1182 & $+8.0419$ & 0.4220 \\
 & $w=-1.1$ & 0.4066 & 1.5117 & 0.1358 & $+7.1580$ & 0.4220 \\
 & CPL & 0.4043 & 1.5181 & 0.1391 & $+7.4509$ & 0.4220
\end{tabular}
\end{ruledtabular}
\end{table*}

For $\zeta<0$, $\gamma>2$ and the matter fixed points are always real. The anchored values satisfy $X_{0,+}>0>X_{0,-}$, so the positive branch is $\mathcal B_+$. The other anchored branch $\mathcal B_-$ reaches a negative-$X$ pole even though the positive branch obeys the comparison bound of Sec.~\ref{sec:theorems}: for the radiation-free $\Lambda$CDM benchmark, the first zero of $\Psi$ occurs at $z_s=0.1392$ for $\zeta=-0.05$ and at $z_s=0.03434$ for $\zeta=-1$. In both cases $\phi\to0$ while $|\dot\phi|$ diverges. Thus the regularity classification refers to the \emph{Riccati/field trajectory}, not to the positivity of the scalar--tensor effective Planck mass, which is discussed separately in Sec.~\ref{sec:pert}. The comparison bound then excludes a positive past-directed Riccati pole as long as the background is regular and satisfies the NEC. Within the radiation-free equations, the positive trajectories approach $X_{\rm m,s}$ as the mathematical matter-dominated limit is taken. This convergence should not be interpreted as a physical prediction at $z\sim10^3$, where radiation must be included.

\subsection{Reconstructed potentials}

For flat FLRW backgrounds, Eq.~\eqref{eq:reconstruction} gives
\begin{equation}
V=-3\zeta H^2\phi^2\mathcal F_V(X;\zeta),
\qquad
\mathcal F_V\equiv1-2X+\frac{X^2}{6\zeta}.
\label{eq:Vfactor}
\end{equation}
At the conformal coupling $\zeta=1/6$, $\gamma=-1$ and the flat Riccati equation factorizes into $dX/dq=(1-X)(A-X)$, with $q=\ln(1+z)$. Since $X=1$ is an exact particular solution for \emph{any} regular flat background, the general branch can be expressed as a single comoving-distance quadrature. Put
\begin{equation}
 \mathcal D_C(z)=\int_0^z\frac{dz'}{E(z')}.
 \label{eq:Dcomoving}
\end{equation}
For $X_0\ne1$, direct integration yields
\begin{align}
 X(z)&=1+\frac{1+z}{E(z)[(X_0-1)^{-1}-\mathcal D_C(z)]},
 \label{eq:Xconformalexact}\\
 \frac{\phi(z)}{\phi_0}&=\frac{E(z)[1-X(z)]}{1-X_0}.
 \label{eq:phiconformalexact}
\end{align}
These formulas also follow from the conformal-stealth solutions of Ref.~\cite{Ayon2013}, expressed here directly in the $X$ variable. Their use is not restricted to a matter-only background. For a flat background with $\Omega_{\rm DE0}=1-\Omega_{\rm m0}-\Omega_{\rm r0}$, anchoring the potential to the dark-energy density at $z=0$ gives $X_{0,\pm}=1\pm\sqrt{\Omega_{\rm DE0}}$. The lower branch, with $X_{0,-}-1<0$, has no denominator zero at positive $z$ because $\mathcal D_C(z)\ge0$. The upper branch has a pole if and only if the distance reaches
\begin{equation}
 \mathcal D_C(z_s)=\Omega_{\rm DE0}^{-1/2}.
 \label{eq:conformalpole}
\end{equation}
The existence and location of the upper-branch pole are therefore determined by an integral of the prescribed expansion history.

For example, taking $\Omega_{\rm m0}=0.3$ and $\Omega_{\rm r0}=9.2\times10^{-5}$ in a flat $\Lambda$CDM background, numerical quadrature of Eqs.~\eqref{eq:Dcomoving}--\eqref{eq:conformalpole} gives $z_s=1.9590$ for the upper branch. The lower branch has $X(3)=0.66549$ and $\phi(3)/\phi_0=1.78472$. At $z=1100$ it gives $\Phi(1100)=300.88$. The radiation-free counterparts are $z_s=1.95816$, $X(3)=0.66533$, $\Phi(3)=1.78442$, and $\Phi(1100)=299.74$, respectively. The agreement is specific to the conformal branch and these benchmark parameters; it does not extend the matter-dominated fixed-point approximation into the radiation era.

At conformal coupling $\mathcal F_V=(1-X)^2$. Defining $W\equiv H\phi(1-X)$, one finds $W=C_W\phi^2$, and hence
\begin{equation}
 V(\phi)=-\frac12C_W^2\phi^4.
 \label{eq:quarticgeneral}
\end{equation}
For an anchored flat solution, including one with radiation, this reduces to
\begin{equation}
 V(\phi)=-\frac{H_0^2\Omega_{\rm DE0}}{2\phi_0^2}\,\phi^4.
 \label{eq:quarticanchored}
\end{equation}
The quartic potential was obtained in Ref.~\cite{Ayon2013}. Its comoving-distance representation here provides a direct test for poles on a prescribed flat background.

For generic couplings evolving toward the stable matter fixed point in the radiation-free benchmark, $\phi\propto(1+z)^{X_{\rm m,s}}$ and $H^2\propto(1+z)^3$. In a radiation-inclusive cosmology these are intermediate matter-era scalings rather than the strict early-time asymptotics. Provided $\mathcal F_V(X_{\rm m,s};\zeta)\neq0$,
\begin{align}
V(\phi)&\propto\phi^{n_{\rm m}},\qquad
n_{\rm m}=\frac{9+\sqrt{73-12/\zeta}}{2},\nonumber\\
\mathcal F_{V,s}&=\frac{(X_{\rm m,s}-1)(4X_{\rm m,s}-3)}{6}.
\label{eq:nm}
\end{align}
For positive coupling the asymptotic potential is negative for $\zeta_{\rm m}\le\zeta<1/6$ and for $\zeta>3/16$, but positive for $1/6<\zeta<3/16$. The special values $\zeta=1/6$ and $3/16$ have vanishing leading coefficient; the former is exactly quartic and the latter is asymptotically quartic at the next nonvanishing order. At $\zeta=3/16$, $X_{\rm m,s}=3/4$ and the approach to this fixed point is $X-X_{\rm m,s}\propto(1+z)^{-3/2}$. Consequently $\mathcal F_V\propto(1+z)^{-3/2}$, $V\propto(1+z)^3$, and $\phi\propto(1+z)^{3/4}$, giving $V\propto\phi^4$. The regular positive branch for $\zeta<0$ has a positive potential in the examples considered here.

For anchored positive-coupling solutions, $V_0=-3\zeta H_0^2\phi_0^2(1-\Omega_{\rm m0})<0$. Hence a regular trajectory with $1/6<\zeta<3/16$ must cross $V=0$ before reaching its positive matter-era asymptote. The roots of $\mathcal F_V$ are
\[
X_{V,\pm}=6\zeta\pm\sqrt{36\zeta^2-6\zeta}.
\]
In the radiation-free $\Lambda$CDM benchmark, the $\zeta=0.175$ lower branch crosses the relevant root at $z\simeq27$ [Fig.~\ref{fig:basins}(c)]. For a comparison model with $\Omega_{\rm r0}=9.2\times10^{-5}$, the radiation-to-matter density ratio at this redshift is $\rho_r/\rho_m\simeq0.0086$; this ratio alone does not determine the shift of the crossing under radiation-inclusive evolution. The positivity argument of Sec.~\ref{sec:theorems} guarantees that the field is monotonic where the total NEC holds, so the potential remains single-valued through the crossing.

% ============================================================================
\section{Stability and observational signatures}\label{sec:pert}
% ============================================================================

\subsection{Scalar normalization and Horndeski description}

Rescaling $\phi\to\lambda\phi$ together with $V(\phi)\to\lambda^2V(\phi/\lambda)$ leaves the background equations unchanged. The otherwise undetermined scalar amplitude may be written as
\begin{equation}
\varepsilon_\phi\equiv\kappa\zeta\phi_0^2,
\qquad
\Phi\equiv\frac{\phi}{\phi_0},
\qquad
\hat F\equiv1-\varepsilon_\phi\Phi^2,
\label{eq:epsphi}
\end{equation}
so that the effective Planck mass is $M_*^2=\mpl^2\hat F$, with $\mpl^2=\kappa^{-1}$. For anchored radiation-free flat solutions, $V_0/\rho_{c,0}=-\varepsilon_\phi(1-\Omega_{\rm m0})$; with radiation, the same expression uses $\Omega_{\rm DE0}$ in place of $1-\Omega_{\rm m0}$.

The action~\eqref{eq:action} is a Horndeski theory~\cite{Horndeski1974,Kobayashi2019} with $G_2=\mathcal X_\phi-V$, $G_4=(\mpl^2-\zeta\phi^2)/2$, and $G_3=G_5=0$, where $\mathcal X_\phi=-\nabla_\mu\phi\nabla^\mu\phi/2$. In the Bellini--Sawicki basis~\cite{Bellini2014},
\begin{align}
\alpha_M&=\frac{2\varepsilon_\phi X\Phi^2}{\hat F}
=-\alpha_B,\qquad \alpha_T=0,\nonumber\\
\alpha_K&=\frac{\varepsilon_\phi X^2\Phi^2}{\zeta\hat F},\nonumber\\
\mathcal D&\equiv\alpha_K+\frac32\alpha_B^2\nonumber\\
&=\frac{\varepsilon_\phi X^2\Phi^2}{\zeta\hat F^2}
\left(\hat F+6\zeta\varepsilon_\phi\Phi^2\right).
\label{eq:alphas}
\end{align}
Because $\varepsilon_\phi/\zeta=\kappa\phi_0^2>0$, $\mathcal D>0$ whenever $\hat F>0$. The scalar principal part is conformally related to that of a minimally coupled canonical field, so wherever the conformal transformation is regular,
\begin{equation}
c_s^2=c_T^2=1,
\qquad
M_*^2>0\;\Longleftrightarrow\;\hat F>0.
\label{eq:highk_stability}
\end{equation}
For regular conformal transformations, $\hat F>0$ excludes high-frequency ghosts and gradient instabilities in the scalar and tensor sectors. The tensor speed satisfies the GW170817 constraint~\cite{GW170817,Creminelli2017,Ezquiaga2017}. Neither statement determines stability at long wavelengths. The gauge-invariant instability found for a different Brans--Dicke stealth background~\cite{FaraoniFranconnet2022} illustrates why a separate coupled perturbation analysis is required; it does not provide a stability verdict for the present solutions. At conformal coupling the coefficient $m_J^2\equiv\zeta R+V_{,\phi\phi}$ satisfies $m_J^2/H^2=2-A-6(1-X)^2$, giving $m_J^2/H_0^2\simeq-2.65$ at $z=0$ for the $\Lambda$CDM benchmark. To gauge the meaning of this negative curvature, one may temporarily neglect metric perturbations. A Fourier mode of the test-field fluctuation then satisfies
\begin{equation}
\delta\ddot\phi_k+3H\delta\dot\phi_k+
\left(\frac{k^2}{a^2}+m_J^2\right)\delta\phi_k=0.
\label{eq:fixedmetric_fluct}
\end{equation}
For $k=0$, numerical integration of this fixed-metric equation on the conformal lower-branch $\Lambda$CDM background from $z=3$ to $z=0$, with $\delta\phi(3)=1$ and $d\delta\phi/d\ln a|_{z=3}=0$, gives $\delta\phi(0)\simeq1.50$. This is only a fixed-metric diagnostic: the gravitationally coupled scalar mode mixes with metric and matter perturbations, and the $k=0$ fluctuation is not by itself a gauge-invariant stability observable. In particular, $m_J^2<0$ alone is not a demonstration of a physical tachyonic instability. A complete horizon-scale analysis remains necessary. The quasistatic expressions below require $k/a\gg H$ and $k^2/a^2\gg|m_J^2|$.

For $\zeta<0$, $\hat F>0$ automatically. For $\zeta>0$, regularity of the Riccati trajectory does \emph{not} ensure $\hat F>0$. If $X>0$ on $[0,z_*]$, Eq.~\eqref{eq:reconstruction} implies that $\Phi$ increases and the positive-effective-Planck-mass requirement becomes\begin{equation}\label{eq:eps_endpoint}0<\varepsilon_\phi<\frac{1}{\Phi^2(z_*)}, \qquad (\zeta>0,\; X>0\text{ on }[0,z_*]).\end{equation} If $X$ is not positive throughout, the general requirement uses the maximum of $\Phi^2$ on the interval rather than its endpoint value. On a past-regular branch with unbounded $\Phi$, any fixed $\varepsilon_\phi>0$ eventually makes $\hat F$ vanish: a globally regular \emph{field profile} is then not a globally viable positive-Planck-mass solution. For the conformal anchored lower branch, Eqs.~\eqref{eq:Xconformalexact}--\eqref{eq:phiconformalexact} give the particularly simple exact formula\begin{equation}\label{eq:Phi_lower_exact}\Phi_-(z)=\frac{1+z}{1+\sqrt{\Omega_{\rm DE0}}\,\mathcal D_C(z)}.\end{equation} In the radiation-inclusive $\Lambda$CDM benchmark, $\Phi_-(1100)=300.88$, so the condition $\hat F>0$ through $z_*=1100$ yields $\varepsilon_\phi<1.10\times10^{-5}$. By contrast, the conditional unscreened Cassini limit~\eqref{eq:cassini} at $\zeta=1/6$ is $\varepsilon_\phi\lesssim3.78\times10^{-5}$: saturating it would give $\varepsilon_\phi\Phi_-^2(1100)\simeq3.42$ and hence $\hat F(1100)<0$. In this example the zero $\hat F=0$ would already occur near $z=589$. Thus the \emph{illustrative viability requirement through recombination} is tighter than Cassini for this branch. It is not a CMB likelihood constraint and makes no claim about nucleosynthesis or other branches.

\subsection{Growth, lensing, and gravitational waves}

In the quasistatic regime the theory has the Brans--Dicke form~\cite{BransDicke1961,EspositoFarese2001}, with
\[
\omega=\frac{\hat F}{4\zeta\varepsilon_\phi\Phi^2}.
\]
Relative to the bare coupling in the action, the effective matter and lensing functions are
\begin{equation}
\mu_{\rm b}=\frac{\hat F+8\zeta\varepsilon_\phi\Phi^2}
{\hat F(\hat F+6\zeta\varepsilon_\phi\Phi^2)},
\qquad
\Sigma_{\rm b}=\frac{1}{\hat F}.
\label{eq:muSigma}
\end{equation}
For $|\varepsilon_\phi|\Phi^2\ll1$, $\mu_{\rm b}-1\simeq(1+2\zeta)\varepsilon_\phi\Phi^2$ and $\Sigma_{\rm b}-1\simeq\varepsilon_\phi\Phi^2$. For phenomenological quantities normalized to the locally measured Newton constant, we use
\begin{equation}
\mu_N(z)=\frac{\mu_{\rm b}(z)}{\mu_{\rm b}(0)},
\qquad
\Sigma_N(z)=\frac{\Sigma_{\rm b}(z)}{\mu_{\rm b}(0)}.
\label{eq:normalized_muSigma}
\end{equation}
Their first-order expansions and the derivation of the local normalization are collected in Appendix~\ref{app:pert}.

The varying effective Planck mass also modifies gravitational-wave friction. The corresponding luminosity distance is~\cite{Lombriser2016,Belgacem2018}
\begin{equation}
\frac{d_L^{\rm GW}(z)}{d_L^{\rm EM}(z)}
=\left[\frac{1-\varepsilon_\phi}{1-\varepsilon_\phi\Phi^2(z)}\right]^{1/2}
\simeq1+\frac{\varepsilon_\phi}{2}\left[\Phi^2(z)-1\right].
\label{eq:dGW}
\end{equation}
Although each observable has a different redshift dependence, its leading correction is proportional to $\varepsilon_\phi$. An observational forecast would require a perturbation calculation and an explicit likelihood; Eqs.~\eqref{eq:muSigma}--\eqref{eq:dGW} alone do not provide one.

\begin{figure*}[t]
\includegraphics[width=\textwidth]{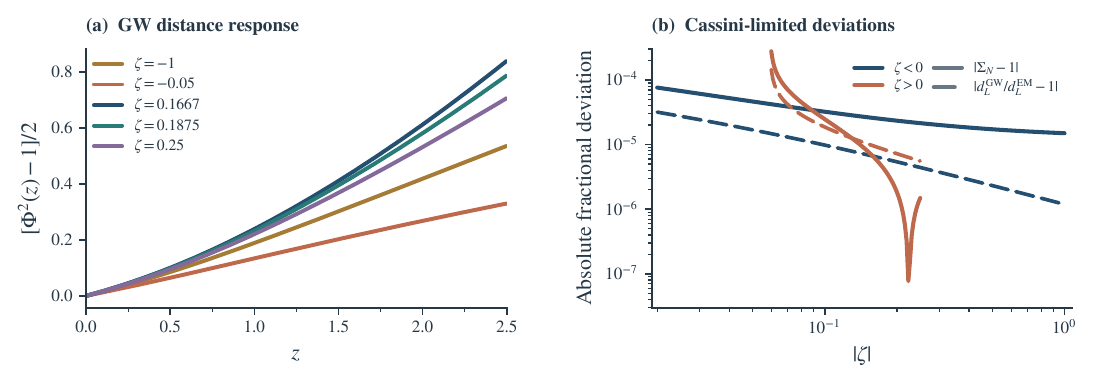}
\caption{(a) Gravitational-wave luminosity-distance response per unit scalar amplitude, $[\Phi^2(z)-1]/2$, for anchored $\Lambda$CDM branches. (b) Cassini-saturated deviations at $z=1$, using the locally normalized lensing function $\Sigma_N$ and discarding parameter values for which the branch does not reach $z=1$ regularly or $M_*^2(z=1)\le0$. These late-time conditional envelopes need not satisfy the stronger requirement $M_*^2>0$ through recombination. The curves were computed for the stated benchmark parameters.}
\label{fig:obs}
\end{figure*}

\subsection{Local constraints}

To apply Solar-System bounds we assume that the scalar remains unscreened and that its local value follows the homogeneous cosmological solution. These assumptions are not implied by $T^{(S)}_{\mu\nu}=0$. The present Brans--Dicke parameter is
\[
\omega_0=\frac{1-\varepsilon_\phi}{4\zeta\varepsilon_\phi},
\]
and therefore $\gamma_{\rm PN}-1=-1/(2+\omega_0)\simeq-4\zeta\varepsilon_\phi$ at small amplitude. The Cassini measurement $\gamma_{\rm PN}-1=(2.1\pm2.3)\times10^{-5}$~\cite{Bertotti2003,Will2014}, together with the model prediction $\gamma_{\rm PN}-1<0$, gives the following approximate two-standard-deviation one-sided limit (using $2.1-2(2.3)\simeq-2.5$ in units of $10^{-5}$):
\begin{equation}
|\varepsilon_\phi|\lesssim\frac{6.3\times10^{-6}}{|\zeta|}.
\label{eq:cassini}
\end{equation}
The same scalar produces a time variation of the locally measured gravitational coupling,
\[
\left.\frac{\dot G_{\rm cav}}{G_{\rm cav}}\right|_0
\simeq-2(1+2\zeta)\varepsilon_\phi X_0H_0,
\]
while lunar laser ranging probes $\dot G/G$ at the $10^{-14}$--$10^{-15}\,\mathrm{yr}^{-1}$ level~\cite{Biskupek2021}. This constraint is not automatically weaker than Cassini: on the conformal lower branch, setting $\varepsilon_\phi=3.78\times10^{-5}$ (the conditional Cassini upper limit), $X_0=1-\sqrt{0.7}$, and illustratively $H_0=70\,\mathrm{km\,s^{-1}\,Mpc^{-1}}$ gives $|\dot G_{\rm cav}/G_{\rm cav}|_0\simeq1.18\times10^{-15}\,\mathrm{yr}^{-1}$, comparable to LLR precision. The stronger background-viability limit through $z=1100$ further lowers this estimate. Neither limit can be applied without assumptions about the local scalar profile.

For the illustrated branches, unscreened local tracking makes the maximal late-time departures small. Environment-dependent screening can change the relation between cosmological and local scalar profiles in other scalar--tensor models~\cite{KhouryWeltman2004}. Such behavior has not been established for the stealth potentials reconstructed here: it would require an explicit inhomogeneous solution and independent local-gravity tests.

% ============================================================================
\section{Conclusions}\label{sec:concl}

For a prescribed FLRW geometry, the vanishing-stress-tensor equations determine the profile of a homogeneous nonminimally coupled scalar and, subsequently, its potential. This reconstruction is independent of an assumed identification between scalar-potential terms and the dark-energy density. In a mixed flat constant-$w$ background, we have shown that imposing such an identification at one epoch does not define an invariant surface of the exact stealth flow. The result excludes a persistent sectorwise interpretation in the stated setting; it does not exclude the stealth configuration itself.

The global analysis separates this compatibility question from branch regularity. In the radiation-free matter--dark-energy benchmark, the Riccati flow admits a hypergeometric representation and a separatrix between regular and singular positive branches. The upper-branch threshold $\zeta_{\rm sep}\simeq0.181$ is obtained for $\Omega_{\rm m0}=0.3$ and depends on the prescribed history, unlike the matter-era discriminant threshold $12/73$. Including radiation preserves the reconstruction equation but changes the asymptotic flow. At conformal coupling, an exact comoving-distance quadrature remains valid with radiation and provides a direct criterion for the upper-branch pole.

The perturbative analysis imposes further restrictions. A positive effective Planck mass gives healthy scalar and tensor propagation in the high-frequency regime. The fixed-metric horizon-scale diagnostic does not establish full coupled stability, nor does background regularity guarantee a consistent perturbative realization of the prescribed CPL sector across $w=-1$. The predicted growth, lensing and gravitational-wave modifications depend on an amplitude absent from the background reconstruction. Under unscreened local tracking, Cassini and lunar laser ranging constrain the amplitude, while positive effective Planck mass through a specified early epoch can impose a tighter, branch-dependent condition; in the conformal benchmark the requirement through $z=1100$ is stronger than Cassini. Extending the analysis to a specified dark-energy perturbation sector and constructing inhomogeneous configurations capable of modifying the local bound remain distinct questions.
\begin{acknowledgments}
J.B. acknowledges financial support from the Secretaría de Ciencia, Humanidades, Tecnología e Innovación (SECIHTI). C.C. acknowledges partial support from the Sistema Nacional de Investigadoras e Investigadores (SNII). G.C. acknowledges financial support from SECIHTI through a postdoctoral fellowship and from SNII. The authors thank Eduardo Cisneros for enlightening discussions.
\end{acknowledgments}

\appendix
% ============================================================================

\section{FLRW reduction of the stealth equations}\label{app:flrw}

For
\[
ds^2=-dt^2+a^2(t)\left[\frac{dr^2}{1-kr^2}+r^2d\Omega^2\right],
\qquad
\phi=\phi(t),
\]
the relevant geometrical quantities are
\begin{align*}
G_{00}&=3\left(H^2+\frac{k}{a^2}\right),\\
G_{ij}&=-\left(2\dot H+3H^2+\frac{k}{a^2}\right)g_{ij},
\end{align*}
and, for any homogeneous $f(t)$,
\begin{align*}
\nabla_0\nabla_0f&=\ddot f,\qquad
\nabla_i\nabla_jf=-Hg_{ij}\dot f,\\
\Box f&=-\ddot f-3H\dot f.
\end{align*}
Applying these identities to $f=\phi^2$, Eq.~\eqref{eq:Tmn} gives
\begin{align}
T^{(S)}_{00}={}&\frac12\dot\phi^2+V
+3\zeta\left(H^2+\frac{k}{a^2}\right)\phi^2\nonumber\\
&+6\zeta H\phi\dot\phi,\nonumber\\
\frac{T^{(S)}_{ij}}{g_{ij}}={}&\frac12\dot\phi^2-V
-\zeta\Bigl[\left(2\dot H+3H^2\right)\phi^2\nonumber\\
&+\frac{k}{a^2}\phi^2
+\frac{d^2\phi^2}{dt^2}+2H\frac{d\phi^2}{dt}\Bigr].
\label{eq:app_components}
\end{align}
Setting the first equation to zero and dividing by $\zeta\phi^2$ gives Eq.~\eqref{eq:Q}. Using $\ddot\phi/\phi=\dot u+u^2$ in the spatial equation and eliminating $Q$ yields Eq.~\eqref{eq:master}. Finally, with $u=-HX$, $\dot H/H^2=-A$, and $k/(a^2H^2)=-\Omega_k$, one recovers Eq.~\eqref{eq:masterz}.

\section{Hypergeometric reduction for constant $w$}\label{app:hyper}

For $\gamma\neq0$, define $\Psi=(\phi/\phi_0)^\gamma$. In $N=\ln a$, the flat constant-$w$ equation is
\begin{equation}
\Psi_{,NN}-\left(\frac52+\frac32w\ODE\right)\Psi_{,N}
-\frac{3\gamma}{2}(1+w\ODE)\Psi=0.
\label{eq:app_PsiN}
\end{equation}
With $r=r_0a^{-3w}$, $\ODE=r/(1+r)$, and $d/dN=-3wr\,d/dr$, Eq.~\eqref{eq:app_PsiN} becomes
\begin{align}
 r^2\Psi_{,rr}&+r\left[1+\frac{5}{6w}
 +\frac{r}{2(1+r)}\right]\Psi_{,r}\nonumber\\
 &-\frac{\gamma}{6w^2}\frac{1+(1+w)r}{1+r}\Psi=0.
\label{eq:Psir}
\end{align}
The Frobenius ansatz $\Psi=r^sF$ gives
\[
s^2+\frac{5}{6w}s-\frac{\gamma}{6w^2}=0,
\qquad
s_\pm=\frac{-5\pm\sqrt{73-12/\zeta}}{12w}.
\]
After imposing the indicial equation,
\[
r(1+r)F_{,rr}+[c_s+(a_s+b_s+1)r]F_{,r}+a_sb_sF=0,
\]
where
\begin{align*}
c_s&=1+\frac{5}{6w}+2s,\\
a_s+b_s&=\frac12+\frac{5}{6w}+2s,\\
a_sb_s&=-s\left(ws+\frac13\right).
\end{align*}
With $x=-r$ this is the standard Gauss equation, giving Eq.~\eqref{eq:hyper}. The present-day normalization fixes
\[
\Psi(r_0)=1,
\qquad
\Psi_{,r}(r_0)=\frac{\gamma X_0}{3wr_0},
\]
and the Riccati variable is recovered as
\[
X=\frac{3wr}{\gamma}\frac{\Psi_{,r}}{\Psi}.
\]
Degenerate Frobenius cases follow by the usual limiting procedure; $\zeta=1/4$ is treated directly at the Riccati level because $\gamma=0$. For present-day arguments $-r_0<-1$ in the benchmark $\Omega_{\rm m0}=0.3$, analytic continuation should be used rather than the defining Taylor series. The Pfaff transformation in the main text provides a real argument $r/(1+r)\in(0,1)$ for $r>0$.

\section{Compatibility algebra}\label{app:compat}

On the sector-preserving surface,
\[
\Om=2X-\frac{X^2}{6\zeta},
\qquad
\ODE=1-\Om.
\]
For a flat constant-$w$ background,
\begin{align*}
\frac{d\Om}{dN}&=3w\Om\ODE,\\
\frac{dX}{dN}&=-A+(1+A)X+\gamma X^2,\\
A&=\frac32(1+w\ODE).
\end{align*}
Differentiating $\delta$ and eliminating $\Om$ gives Eq.~\eqref{eq:flow}, with the expanded cubic
\begin{align}
\mathcal N_{w,\zeta}(X)={}&
(6w\zeta-w+8\zeta-2)X^3
\nonumber\\
&+\bigl[-72w\zeta^2+18w\zeta
 -48\zeta^2+22\zeta\bigr]X^2\nonumber\\
&-\bigl[36w\zeta^2+6w\zeta
 +60\zeta^2+6\zeta\bigr]X\nonumber\\
&+36\zeta^2(1+w).
\label{eq:Nexp}
\end{align}
This expression is algebraically equivalent to the factorized form in Eq.~\eqref{eq:Npoly}. At $w=-1$ and $\zeta=1/6$ it reduces to
\[
\mathcal N_{-1,1/6}(X)=-\frac23X(X-1)^2,
\]
so strict compatibility occurs only at the matter- and dark-energy-dominated endpoints, not throughout the mixed $\Lambda$CDM epoch.

The factorized form also yields the instantaneous equation of state required by exact sector preservation,
\begin{equation}
w_{\rm sp}(X;\zeta)=
-\frac{2(X-6\zeta)[(4\zeta-1)X^2+5\zeta X-3\zeta]}
{[X^2-12\zeta X+6\zeta][(6\zeta-1)X+6\zeta]},
\label{eq:app_wsp}
\end{equation}
whenever the denominator is nonzero. For a time-dependent background such as CPL, strict sector preservation therefore imposes a functional relation between the prescribed $w(z)$ and the reconstructed trajectory $X(z)$.

\section{Perturbative and local-gravity relations}\label{app:pert}

For convenience, we collect here the first-order expressions used in Sec.~\ref{sec:pert}. Expanding Eq.~\eqref{eq:muSigma} at $|\varepsilon_\phi|\Phi^2\ll1$ gives
\[
\mu_{\rm b}-1\simeq(1+2\zeta)\varepsilon_\phi\Phi^2,
\qquad
\Sigma_{\rm b}-1\simeq\varepsilon_\phi\Phi^2.
\]
Normalizing to the present Cavendish coupling,
\begin{align}
\mu_N(z)-1&\simeq(1+2\zeta)\varepsilon_\phi[\Phi^2(z)-1],
\nonumber\\
\Sigma_N(z)-1&\simeq\varepsilon_\phi[\Phi^2(z)-(1+2\zeta)].
\label{eq:app_muSigmaN}
\end{align}
The present Brans--Dicke parameter is
\[
\omega_0=\frac{1-\varepsilon_\phi}{4\zeta\varepsilon_\phi},
\]
which implies
\[
\gamma_{\rm PN}-1=-\frac{1}{2+\omega_0}
\simeq-4\zeta\varepsilon_\phi
=-4\kappa\zeta^2\phi_0^2.
\]
Applying the lower end of the approximately $2\sigma$ Cassini interval, with the sign $\gamma_{\rm PN}-1<0$ imposed by the model, gives Eq.~\eqref{eq:cassini}. The leading time variation of the locally measured coupling follows from the same quasistatic expression,
\[
\left.\frac{\dot G_{\rm cav}}{G_{\rm cav}}\right|_0
\simeq-2(1+2\zeta)\varepsilon_\phi X_0H_0.
\]
These relations assume that the scalar is unscreened locally and that its Solar-System value follows the cosmological background.

% ============================================================================
% ============================================================================

\bibliographystyle{apsrev4-2}
\bibliography{references}

\end{document}